\documentclass[aps,twocolumn,prl,superscript,floatfix,superscriptaddress,10pt]{revtex4-2}

\usepackage{epsfig}
\usepackage{graphicx}
\usepackage{dcolumn}
\usepackage{bm}
\usepackage{amsfonts}
\usepackage{amsmath}
\usepackage{amssymb}
\usepackage{mathrsfs}
\usepackage{natbib}
\usepackage{multirow}

\usepackage[colorlinks=true,%
            linkcolor=blue,%
            urlcolor=blue,%
            citecolor=blue,%
            filecolor=blue,%
            bookmarksopen=true,%
            pdfauthor={M.Figueroa},%
            pdftitle={QR},%
            pdfsubject={QR_manuscript},%
            pdfpagemode=UseOutlines]{hyperref}

\usepackage[T1]{fontenc} 
\usepackage{lmodern}
\usepackage{xcolor}

\begin{document}

\title{Exceptional magic angles in non-Hermitian  twisted bilayer graphene}
\author{Juan Pablo Esparza}
\affiliation{Departamento de F\'isica, Universidad T\'ecnica Federico Santa Mar\'ia, Casilla 110, Valpara\'iso, Chile}
\affiliation{Instituto de F\'isica, Pontificia Universidad Cat\'olica de Valpara\'iso, Avenida Universidad 331, Curauma, Valpara\'iso, Chile}

\author{Vladimir Juri\v ci\'c}\thanks{Corresponding author:vladimir.juricic@usm.cl}
\affiliation{Departamento de F\'isica, Universidad T\'ecnica Federico Santa Mar\'ia, Casilla 110, Valpara\'iso, Chile}

\begin{abstract}
Twisted bilayer graphene (TBG) features strongly correlated and topological phases due to its flat bands emerging near the magic angle. However, the effects of the non-Hermiticity, arising from the coupling to the environment and dissipation, have remained unexplored. We here develop a simple non-Hermitian (NH) version of twisted bilayer graphene (TBG) by considering relative twisting of two NH graphene monolayers with non-Hermiticity encoded in the imbalance of in-plane nearest-neighbor hopping amplitudes. Remarkably, by generalizing the Bistritzer-MacDonald approach to NH systems, we discover \emph{exceptional magic angles} where the band structure changes  from purely real to purely imaginary thus featuring flat bands with infinite lifetime. Between them, the bands remain flattened, and a \emph{Hermitian magic angle} emerges at which the imaginary part of energy is maximal, and corresponds to the usual magic angle in non-dissipative, purely Hermitian TBG. We  propose an optical lattice setup with gain and loss  where our theoretical predictions can be verified. These results suggest the robustness of the flat bands in open systems, paving the way for the further studies on the interplay of dissipative effects, electronic topology, and interactions in such NH moiré bands.
\end{abstract}
\maketitle

\emph{Introduction.} Twisted bilayer graphene (TBG) has emerged as a fascinating platform for exploring a wide range of strongly correlated insulating and superconducting quantum phases since the first experimental reports~\cite{cao2018correlated,cao2018unconventional}. The origin of such rich phenomenology lies in the presence of nearly flat electronic bands~\cite{suarez2010flat,bistritzer2011moire,dosSantos2007,Pankratov-PRB2010}, where electronic interactions become dominant, giving rise to a landscape of exotic quantum phases, such as topological integer~\cite{choi2019electronic,Zhang2019Nearly,nuckolls2020strongly,choi2021correlation,das2021symmetry,wu2021chern,saito2021hofstadter,park2021flavour} and fractional Chern insulators~\cite{xie2021fractional}, ferromagnetic states~\cite{sharpe2019emergent,saito2021hofstadter,lin2022spin}, correlated nematics~\cite{Cao2021Nematicity,rubio2022moire} and superconductors~\cite{Yankowitz2019Tuning,saito2020independent,oh2021evidence} that can be controlled by external parameters, e.g. by doping and applied magnetic fields. These flattened bands are particularly pronounced near the so-called magic angles~\cite{suarez2010flat,bistritzer2011moire}, where the Fermi velocity of the Dirac quasiparticles approaches zero, with the flatness persisting throughout the moir\' e Brillouin zone (mBZ) of the TBG, due to the subtle interplay between the band topology and symmetries of the moir\' e lattice~\cite{Senthil-PRX2018,kang2018symmetry,Kruchkov2019,Bernevig-PRL2019,bernevig2021twisted1,sheffer2023symmetries}.

Flat bands can be affected by dissipation and coupling to the environment, leading to, for instance, finite lifetimes of electronic states~\cite{Leykam-PRB2017,Zyuzin-PRB2018,Kozzi2024lifetime}, and they may even emerge as a result of dissipation~\cite{talkington2022dissipation}. To study these non-Hermitian (NH) phenomena, biorthogonal quantum mechanics is employed, where the Hermitian constraint on the Hamiltonian operator is lifted, and the dynamics is governed by an NH Hamiltonian-like operator~\cite{brody2013biorthogonal}. The exploration of NH effects has been extensive in topological quantum matter~\cite{Ueda-PRX2018,torres2019perspective,Bergholtz-2021,ding2022non,yang2024homotopy}, with pure non-Hermiticity-driven features, such as band structures with exceptional points where the eigenvectors and eigenstates coalesce~\cite{Lee-2016,Hu-exceptional-2017,sato-exceptional-2019,Murakami-BBC-2019}, unusual bulk-boundary correspondence~\cite{Kunst-PRL2018,Shen-Fu-2018,song-2019,Slager-2020} and the NH skin effect~\cite{Foa-2018,yao-wang-PRL2018,yao-wang-nh-chern-2018,thomale-2019}. This endeavor  has recently extended to interacting Dirac quantum  materials~\cite{jurivcic2024yukawa,murshed2024quantum,murshed2024yukawalorentz,Xiang-2024}, opening a new research front in their exploration. Nevertheless, the impact of NH effects on the TBG flat bands remains largely unexplored. It  is expected  that such a synergy of the non-Hermitian and  flat-band physics should provide valuable insights into the stability and potential formation of novel states of quantum matter tunable via non-Hermiticity, such as NH Mott insulators~\cite{Liu2020Mott} and superconductors~\cite{kawabata2019topological}. We here show that the main effect of non-Hermiticity on  flat bands in TBG is the emergence of the new kind of magic angle, \emph{the exceptional magic angle} (EMA), at which the spectrum of the effective Hamiltonian collapses to zero energy, with both the real and imaginary components of the Dirac fermion velocity vanishing, and the bands thus flattened.

\begin{figure}[h]
    \centering
    \includegraphics[width=\linewidth]{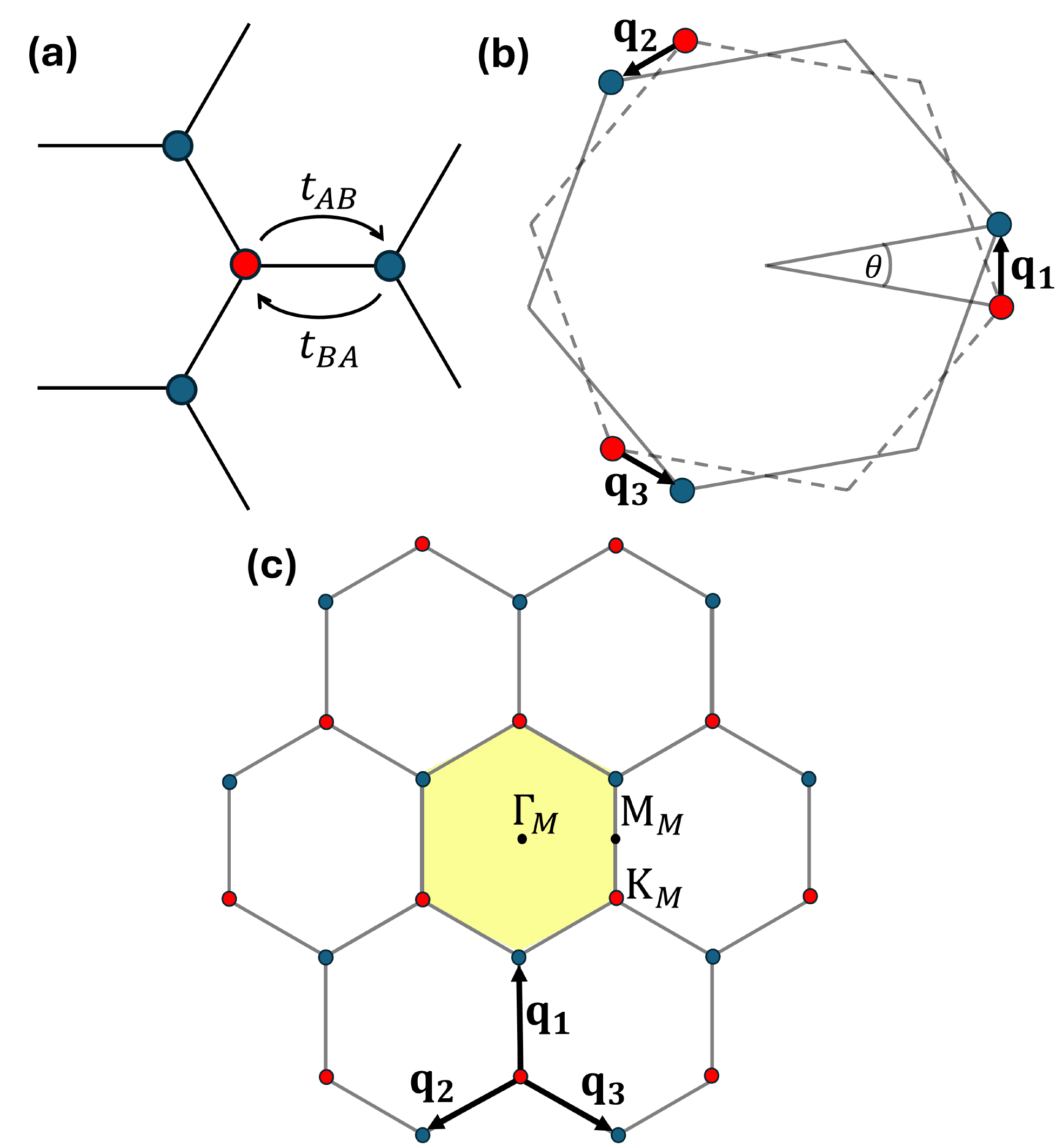}
    \caption{Lattice model and geometry of non-Hermitian (NH) twisted bilayer graphene (TBG). (a) The non-Hermiticity arises from an asymmetric (non-reciprocal)  nearest-neighbor (inter-sublattice) hopping in the graphene plane, $t_{AB}\neq t_{BA}$, with  $\delta t =(t_{AB}-t_{BA})/2\neq 0$ as the corresponding tuning parameter. In the Hermitian case, $\delta t=0$, $\bar{t}=(t_{AB}+t_{BA})/2$ becomes the effective hopping amplitude. (b) The relative angle $\theta$ sets the size of the momenta $|\mathbf{q}_i|=2 k_D \sin(\theta/2)\equiv k_\theta$ generating the momentum space corresponding to the moiré lattice of the TBG, where $k_D$ is the magnitude of the $K$-point vector in the Brillouin zone (BZ) of the monolayer graphene. (c) The reciprocal lattice of TBG spanned by $\mathbf{q_i}$ representing the difference between the $K$-point vectors in the two layers. Red and blue dots represent  the Dirac points in the bottom and top graphene layer, respectively. The moiré BZ is shown in yellow   with its high-symmetry points.} 
    \label{fig:set_up}
\end{figure}

\emph{Minimal model.} As a starting point, we develop a minimal NH version of TBG, by  considering  a simple NH extension of the low-energy Dirac Hamiltonian for  monolayer graphene. The latter  is obtained from the usual  tight-binding model for graphene but with  an asymmetry imposed  between nearest-neighbor hopping amplitudes from the sublattice $A$ to $B$ ($t_{AB}$) and  $B$ to $A$ ($t_{BA}$)~\cite{jurivcic2024yukawa}:
\begin{equation}
   h_\mathbf{k}=(v_{\text{H}}+v_{\text{NH}}\sigma_3)\boldsymbol{\sigma}\cdot\mathbf{k}.
\label{eq:monolayer_hamiltonian}
\end{equation}
Here, $v_{\text{H}}={3\bar{t}}/{2a}$ is the Fermi velocity associated with the mean-value of nearest-neighbor hopping amplitudes, $\bar{t}=(t_{AB}+t_{BA})/2$, and $v_{\text{NH}}=3\delta t/2a$, with $\delta t =(t_{AB}-t_{BA})/2\neq 0$ as the hopping imbalance, therefore yielding the NH part of the Hamiltonian, see Fig.~\ref{fig:set_up}(a). $a$ is the nearest-neighbor distance,  ${\bf k}$ is deviation from the momentum at the band-touching, $K_\pm$ points in the hexagonal Brillouin zone, and ${\sigma_\alpha}$, $\alpha=0,1,2,3$, are the Pauli matrices acting in the sublattice space. Such an NH Hamiltonian features a linear dispersion,
\begin{equation}\label{eq:VF}
\epsilon_{\bf k}=v_F k,\,\,\,{\rm with} \,\,\,v_F=v_{\rm H}\sqrt{1-\beta^2},
\end{equation}
 as the effective Fermi velocity,   $\beta=v_{\rm NH}/v_{\rm H}$, and $v_{\rm H}>0$.  The spectrum is therefore purely real (imaginary) for $|\beta|<1$ ($|\beta|>1$), with exceptional points realized for $|\beta|=1$. The corresponding density of states is $\rho(\epsilon)=\frac{N_f}{2\pi v_F^2}\{|\epsilon|\Theta(1-|\beta|)+{\Lambda^2}\delta(\epsilon)\Theta(|\beta|-1)\}$, with $ N_f$ as the number of two-component flavors, $\Theta(x)$ is the Heaviside step function, and $\Lambda$ is the cutoff energy scale up to which the linear dispersion in Eq.~\eqref{eq:VF} holds (see Sec.~S1 of the Supplementary Information (SI)~\cite{SI}). Notice the pronounced zero-energy peak which  arises only when the spectrum is purely imaginary, signaling  band-flattening that emerges at the exceptional points.  Crucially,  non-spatial unitary and antiunitary particle-hole symmetries (PHS)~\cite{Bernard2002} protect this NH Dirac Hamiltonian:  ($i$) unitary PHS, $\mathcal{S}h_{\bf k}\mathcal{S}^{-1}=-h_{\bf k}$, with $\mathcal{S}=\sigma_3$; ($ii$) anti-unitary PHS,  $\mathcal{C}_- h_{\bf k}^\dagger \mathcal{C}_-^{-1}=-h_{\bf k}$, with $\mathcal{C}_-=\sigma_1\mathcal{K}$; and ($iii$) conjugate time-reversal symmetry (${\rm TRS}^\dagger$), $\mathcal{C}_+ h_{\bf k}^\dagger \mathcal{C}_+^{-1}=h_{\bf k}$, with $\mathcal{C}_+=\sigma_2\mathcal{K}$, where $\mathcal{K}$ is the complex conjugation.
 Finally, the composite $C_2{\mathcal{T}}$ symmetry, with $C_2$ (${\mathcal{T}}$) as the two-fold rotation (time-reversal), represented by $\sigma_1$ ($\mathcal{K}$), maps 
 $H({\bf k})$ into $H^\dagger({\bf k})$,
 $C_2 {\mathcal{T}} H({\bf k}) C_2{\mathcal{T}}=H^\dagger({\bf k})$.

To construct the effective continuum model for the NH TBG forming a moir\' e lattice at a small twist angle ($\theta$), following the Bistritzer-MacDonald (BM) approach~\cite{bistritzer2011moire}, we only consider the low-energy processes where the  electrons close to one of the $K$-points in one (bottom) layer  hybridize with  any of the three equivalent Dirac points in the other (top) layer through  Hermitian hoppings. These processes are  chosen to be Hermitian since they describe van der Waals interlayer hybridization  in quasi-two-dimensional heterostructures, such as TBG, while  possible NH terms can be generated indirectly by the in-plane NH terms and are therefore expected to be suppressed. The resulting  low-energy Hamiltonian of the bilayer system is 
\begin{equation}
\mathcal{H}_\mathbf{k}=\begin{pmatrix}
        h_\mathbf{k} & T_1 & T_2 & T_3\\ T_1 & h_{\mathbf{k}_1} & 0 & 0 \\ T_2 & 0 & h_{\mathbf{k}_2} & 0 \\ T_3 & 0 & 0 & h_{\mathbf{k}_3}
    \end{pmatrix},
\label{eq:bilayer_hamiltonian}
\end{equation}
with $h_{\bf k}$ in Eq.~\eqref{eq:monolayer_hamiltonian}. 
Here,  we define $\mathbf{k}_i=\mathbf{k}+\mathbf{q}_i$ for the three possible (small) momenta  transferred through the hybridization processes [Fig.~\ref{fig:set_up}(b)], $|\mathbf{q}_i|=2K_D \sin(\theta/2)$ as the momentum difference between the $K$-points in the two layers, and twist-dependent contribution in the NH Dirac terms, $h_{{\bf k}_i}$, neglected. The hopping matrices read~\cite{bistritzer2011moire}
\begin{equation*}
T_n=w_0\sigma_0+w_1\cos\left[(n-1)\phi\right]\sigma_1+w_1\sin\left[(n-1)\phi\right]\sigma_2,
\end{equation*}
where $w_0$ ($w_1$) stands for the interlayer hopping amplitude between the same (different) sublattices and $\phi=2\pi/3$ is the angle between carbon bonds.


The emergence of EMAs can be understood through general principles concerning the effects of non-Hermiticity. Topology and symmetry dictate that at small twisting angles and weak interlayer hybridization, the effective Hamiltonian retains a Dirac form in the Hermitian case~\cite{Senthil-PRX2018,sheffer2023symmetries}, which is also the case with non-Hermiticity, as we explicitly show [Eq.~\eqref{eq:Ham-NH-TBG}]. In the NH Dirac Hamiltonian, the effective Fermi velocity is determined by the difference between the squares of the Hermitian ($v_{\rm H}$) and NH ($v_{\rm NH}$) velocities. For maximal kinetic energy quenching via twisting, the Hermitian velocity should decrease, analogously as in purely Hermitian TBG, while the NH velocity should  simultaneously increase. This behavior occurs when the spectrum is real ($|\beta|<1$), resulting in the absolute values of these velocities meeting and producing an EMA in the NH TBG Dirac Hamiltonian. Additionally, the effective Fermi velocity’s non-monotonic behavior with the twist angle results in a minimum at the Hermitian magic angle (HMA). Thus, at least two EMAs are expected, with an HMA in between, as we found in NH TBG.

To show that the effective NH Hamiltonian for the twisted system retains the Dirac form, we first isolate the zero-energy part in Eq. \eqref{eq:bilayer_hamiltonian}:
\begin{equation}
\mathcal{H}_\mathbf{k}=\mathcal{H}^{(0)}+\mathcal{H}^{(1)}.
\label{eq:perturation_hamiltonian}
\end{equation}
Here, $\mathcal{H}^{(0)}=\mathcal{H}_{\mathbf{k}=0}$, and $\mathcal{H}^{(1)}$ is a perturbation. Then, we find the zero-energy  eigenstates for $\mathcal{H}^{(0)}$, denoted as $\Psi_{\alpha}^T=(\psi_0,\psi_1,\psi_2,\psi_3)_\alpha$, where $\alpha=R/L$ grades the right and left eigenstates, forming the biorthogonal basis, as dictated by the non-Hermiticity~\cite{brody2013biorthogonal} (Sec.~S2 of SI~\cite{SI}). The eigenvalue problem for the right solution then reduces to the conditions on the components of the zero-energy mode in the top layer ($j=1,2,3$)
\begin{align}
    h_0\psi_0+\sum_j T_j\psi_j&=0,\label{eq:NH-ZEM-1}\\
    T_j \psi_0+h_j\psi_j&=0.\label{eq:NH-ZEM-2}
\end{align}
Here, $\psi_0$ is the zero-mode for the in-plane NH Dirac Hamiltonian in the bottom layer.
Remarkably, in spite of the non-Hermiticity, the components in the top layer   can be written solely in terms of the zero-energy  state of the original (untwisted) NH Dirac Hamiltonian in the bottom layer, up to a trivial shift, yielding
\begin{equation}
    \left(h_0-i 6 \alpha_0 \alpha_1 v_{\text{NH}} k_\theta\right)\psi_0=0,
\label{eq:eigenvalue_problem}
\end{equation}
 with $\alpha_i=w_i/(v_F k_\theta)$, which moves the band touching point to a finite imaginary energy in a generic continuum model. However, since the shift is trivial (proportional to the unity matrix),  the form of the basis does not change, and therefore we can  use it to construct the effective NH Hamiltonian, starting  from $\mathcal{H}^{(1)}$ via perturbation theory for NH systems~\cite{sternheim1972non}. Notice that when either of the tunnelings between the layers vanishes, the zero-energy state is not shifted even when  $v_{\text{NH}}\neq0$. In particular, in the chiral NH model, $w_0=0$, the Dirac point remains at zero energy. However, even in this case,  this  does not imply that the zero-energy states keep the same form as in the Hermitian model due to the non-Hermiticity of the in-plane  Hamiltonian. The right and left  zero-energy modes in the NH TBG thus read,
\begin{equation}
    \Psi_R=\begin{pmatrix}
        \psi_{0,R} \\ -h_j^{-1}T_j \psi_{0,R}
    \end{pmatrix},\,\, \Psi_L=\begin{pmatrix}
        \psi_{0,L} \\ -(h_j^\dagger)^{-1} T_j \psi_{0,L}
    \end{pmatrix},
\label{eq:zero_energy_eigenstates}
\end{equation}
with the norm of the  zero-energy state 
\begin{equation}
\Psi_L^{\dagger}\Psi_R=1+3(\alpha_0^2+\alpha_1^2).
\label{eq:normalization}
\end{equation}
Notice that this is the same result as in the Hermitian model~\cite{bernevig2021twisted}, with the Fermi velocity replaced by the effective one in the NH monolayer graphene, $v_F$, given by Eq.~\eqref{eq:VF}, and in agreement with the BM result for the Hermitian model in the isotropic limit $w_0=w_1=w$~\cite{bistritzer2011moire}.

\begin{figure}[t]
    \centering
\includegraphics[width=\linewidth]{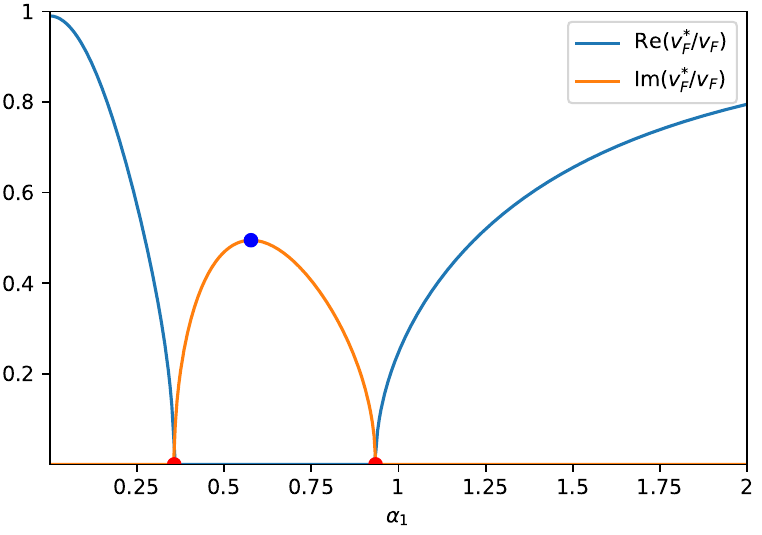}
    \caption{The renormalized Fermi velocity $v_F^*=\sqrt{(v_H^*)^2-(v_{NH}^*)^2}$ in the continuum non-Hermitian Dirac model for twisted bilayer graphene,  Eq.~\eqref{eq:renormalized_velocities}. Here, the  dimensionless interlayer tunneling parameter $\alpha_1=\frac{w_1}{v_F k_\theta}$,  and  $\alpha_0=0$. The hermitian magic angle (blue dot) is given by the maximum value of the imaginary part, at $\alpha_1=\frac{1}{\sqrt{3}}\approx 0.57$. At the exceptional magic angles (red dots) the effective Fermi velocity vanishes. We fix $v_F=1$  ($v_{\rm NH}=0.5$) for the effective Fermi (non-Hermitian) velocity of the graphene monolayer. See Sec.~S5 of SI~\cite{SI} for additional plots.}
    \label{fig:effective_vf}
\end{figure}

We now construct the continuum Hamiltonian for NH TBG using  this biorthogonal basis of the NH quasi-zero-modes in the mBZ by projecting the Dirac (momentum-dependent) piece of the total Hamiltonian [Eq.~\eqref{eq:perturation_hamiltonian}] onto this basis, yielding 
\begin{equation}
\mathcal{H}^{(1)}=\boldsymbol{1}_4\otimes(v_{\text{H}}+v_{\text{NH}}\sigma_3)\boldsymbol{\sigma}\cdot\mathbf{k},
\end{equation}
with matrix elements of the projector 
\begin{align*}
    \langle i |\mathcal{H}^{(1)}|j\rangle=\frac{\Psi_L^{(i)\dagger}\mathcal{H}^{(1)}\Psi_R^{(j)}}{\Psi_L^{(i)\dagger}\Psi_R^{(j)}},
\end{align*}
and we use the indices $i,j=\pm$ to denote the two zero-energy states of the in-plane Dirac Hamiltonian,  $h_\mathbf{k}$. The effective continuum Hamiltonian then reads
\begin{equation}\label{eq:Ham-NH-TBG}
   \mathcal{H}^{\rm eff}_\mathbf{k}=(v_{\text{H}}^*+v_{\text{NH}}^*\sigma_3)\boldsymbol{\sigma}\cdot\mathbf{k},
 \end{equation}
with renormalized velocities
\begin{equation}
    \frac{v_{\rm H}^*}{v_{\rm H}}=\frac{1-3\alpha_1^2}{1+3(\alpha_0^2+\alpha_1^2)},\,\,\frac{v_{\rm NH}^*}{v_{\rm NH}}=\frac{1+3\alpha_1^2}{1+3(\alpha_0^2+\alpha_1^2)},
\label{eq:renormalized_velocities}
\end{equation}
and ${\bf k}$ is the momentum measured from the $K_M$ point in the mBZ [Fig.~\ref{fig:set_up}(c)]. See Sec.~S3 of the SI~\cite{SI} for the details of the derivation. Therefore, the continuum Hamiltonian of the NH TBG takes an NH Dirac form, identical to the in-plane one, with the Hermitian (non-Hermitian) velocity that decreases (increases) due to the interlayer hybridization. Furthermore,  the Dirac points are protected by the same unitary and anti-unitary symmetries as in the monolayer graphene, with the Berry phase of the NH Dirac Hamiltonian quantized to $\pi$ (Sec.~S4 of the SI~\cite{SI}). Furthermore, the effective NH Dirac Hamiltonian in Eq.~\eqref{eq:Ham-NH-TBG} captures salient features of different lattice models of   NH TBG (Sec.~S6 of the SI~\cite{SI}). 
Its form  has rather nontrivial consequences for the band structure of  NH TBG, particularly, the existence of the special twist angles with pronounced band flattening, which we discuss next.

\begin{figure}[t]
    \centering
    \includegraphics[width=\linewidth]{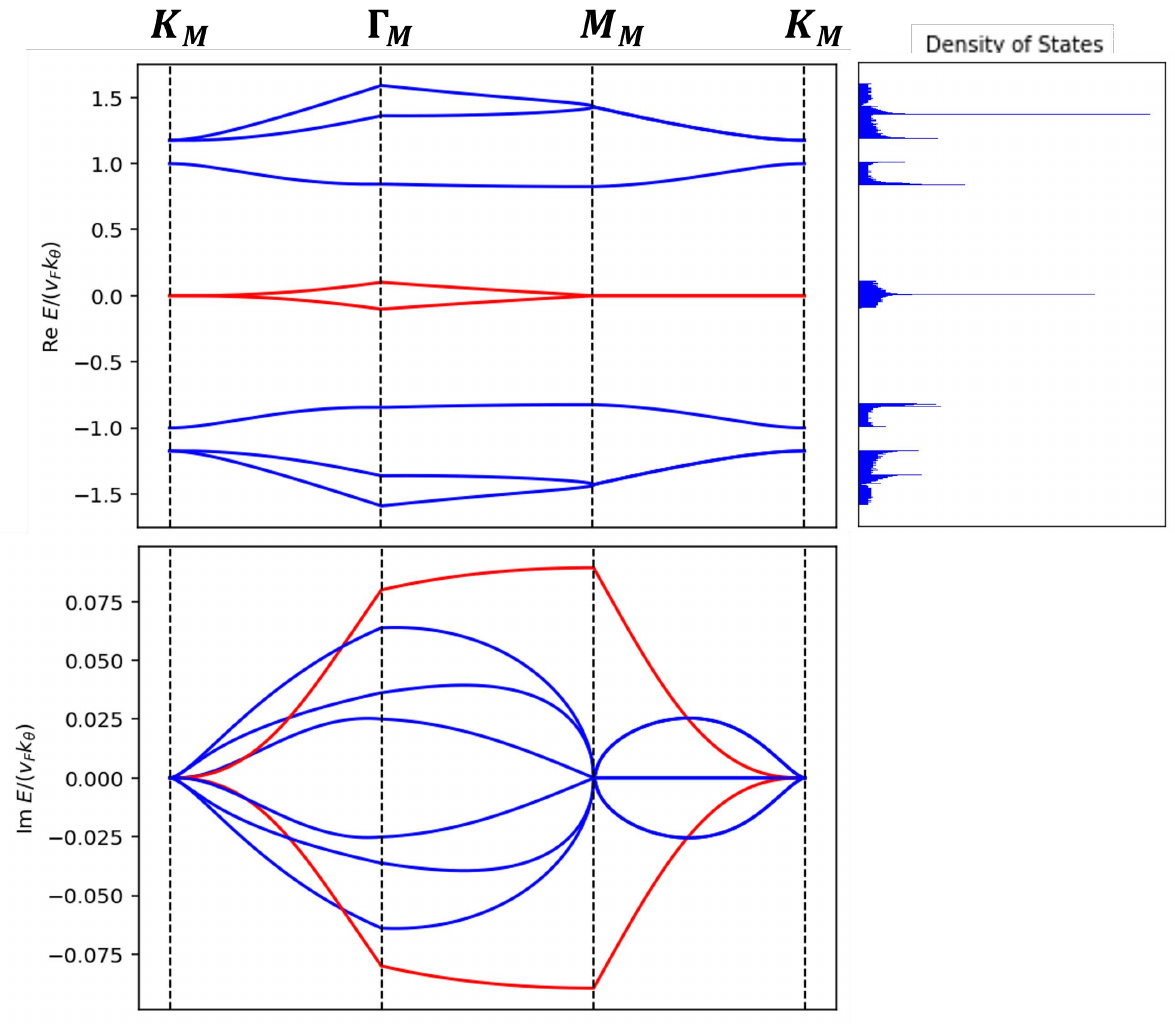}
    \caption{Real (top) and imaginary (bottom) parts of the energy for the bands in the 8-band model in Eq.~\eqref{eq:bilayer_hamiltonian} for $w_0=0$ at the first exceptional magic angle given by  $\alpha_{\rm EMA}^{(+)}$ in  Eq.~\eqref{eq:EMAs}. The flat bands (red)  exhibit near-zero real and imaginary parts in the vicinity of the $K_M$ point. We use the values of the in-plane velocities  $v_F=1$ and  $v_{\rm NH}=0.5$. Notice that the  density of states (DOS) exhibits a sharp peak at zero energy signaling the band flattening feature,  consistent with the form of the DOS, see discussion after Eq.~\eqref{eq:VF} and Sec.~S1 of the SI. Notice that the peak is smeared near the $K_M$ point due to a finite imaginary part of the energy   yielding a  finite lifetime.} 
    \label{fig:1stEMA}
\end{figure}

\emph{Exceptional magic angles (EMAs).} Notice first that the band structure of the NH TBG depends linearly on the \emph{renormalized Fermi velocity} $v_{\rm F}^*=\sqrt{(v_{\rm H}^*)^2-(v_{\rm NH}^*)^2}$, and therefore its vanishing implies the appearance of the exceptional points for certain twist angles, \emph{the EMAs},
\begin{equation}\label{eq:EMAs}
    \alpha^{(s)}_{\rm EMA}=\sqrt{\frac{1-s\beta}{3(1+s\beta)}}
\end{equation}
where $s=+$ ($s=-$) corresponds to the first (second) EMA. Therefore, the EMAs reduce to the usual magic angle in the purely Hermitian case  $\beta\to0$. However, for a finite (inplane) non-Hermiticity $|\beta|<1$, two EMAs appear close to the angle at which the Hermitian piece of the Hamiltonian vanishes, see  Fig.~\ref{fig:effective_vf}. Their appearance is a direct consequence of the form of the effective  NH TBG Dirac Hamiltonian with the spectrum controlled by the effective Fermi velocity, $v^*_F$. Their emergence  is also consistent with the expected maximal depletion of the Fermi velocity by twisting. Finally, the bands remain at zero real energy for the angles between the two EMAs, and not only at the magic angle as in the purely Hermitian TBG, with non-Hermiticity therefore enriching the behavior of the flat bands in TBG.

\emph{Hermitian magic angle (HMA).}  Within the flat band regime, a particularly important twist angle is for which the Hermitian velocity, $v_{\rm H}^*$, vanishes.
Notably, such a {HMA} features a maximum effective  NH velocity, $v_{\rm NH}^*$, and its value coincides with that of the magic angle in the Hermitian TBG, independently of the tunneling $w_0$, see Eq.~\eqref{eq:renormalized_velocities}.
However, for  $w_0\neq0$, therefore breaking PHS, the NH features are   particularly pronounced  through the shift of the Dirac point to a finite imaginary energy,  Eq.~\eqref{eq:eigenvalue_problem}, corroborated by the non-vanishing  $v_{\rm NH}^*$.

\emph{8-band continuum model.} To further support our scenario of the emergent NH flat bands, we numerically analyze the 8-band continuum model, given by Hamiltonian in Eq.~\eqref{eq:bilayer_hamiltonian}.  The corresponding band structure along the path in the mBZ $K_M\to\Gamma_M\to M_M\to K_M$  for the chiral model ($w_0=0$) at the first EMA  is displayed in Fig.~\ref{fig:1stEMA}. We first notice the doubly degenerate flattened  band along the high symmetry $M_M-K_M$ line,  which emerges from the predicted flat band exactly at the K valley, and therefore extends  in the mBZ, as also corroborated by the density of states. Furthermore, the imaginary part of the energy  approaches zero close to the $K_M$-point, as governed by the effective NH Dirac model, but, in fact, the flattening of the imaginary part of the energy also extends through the mBZ. The band structure plots at the HMA and the second EMA exhibit flat bands in real energy, and the imaginary part close to the HMA is an order of magnitude larger than at the EMAs,  see  Sec.~S5 of SI~\cite{SI}, in qualitative  agreement with our theoretical findings.

 \begin{figure}[t!]
    \centering
    \includegraphics[width=\linewidth]{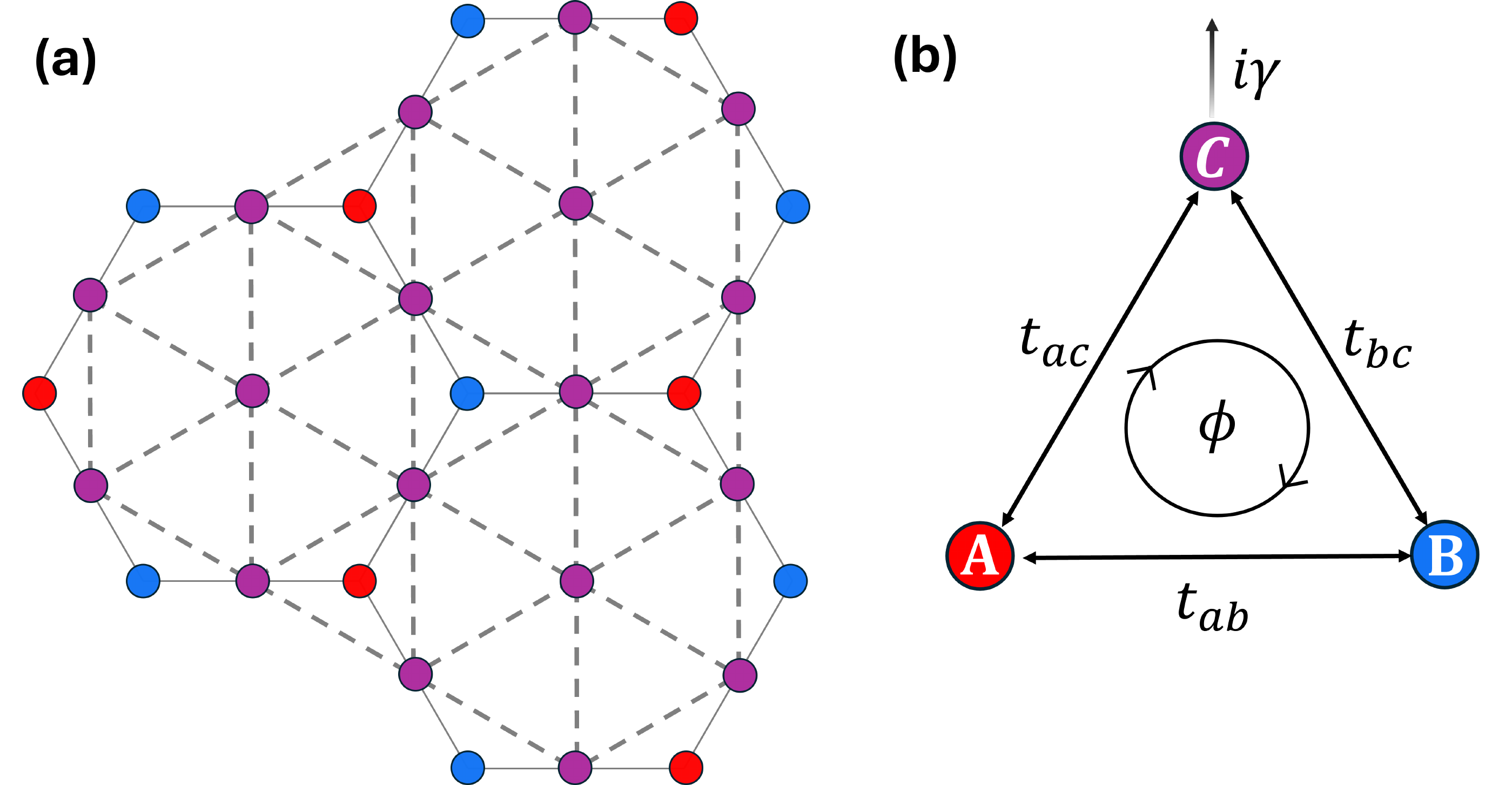}
    \caption{Experimental proposal for realization of non-Hermitian (NH) twisted bilayer graphene (TBG). Dissipation and non-reciprocal hoppings are introduced in the optical honeycomb lattice by means of an auxiliary lattice coupled to an external reservoir. Such two Bernal-stacked lattices then hybridize [Eq.~\eqref{eq:bilayer_hamiltonian}] to realize NH TBG.  {(a)} A triangular auxiliary lattice (purple dots) overlaid on a honeycomb lattice (red and blue dots). {(b)} The tunneling processes between sites of the original lattice ($A$ and $B$) and the auxiliary one ($C$). The parameter $\gamma$ is the coupling with the external reservoir  and $\phi$ is the synthetic magnetic flux~\cite{NH-SSH-experiment}.}
    \label{fig:experimental_proposal}
\end{figure}

{\emph{Experimental proposal.} 
 Concerning  the possible experimental implementation of NH TBG, we note that Hermitian TBG with tunable interlayer tunneling can be achieved by either coupling two overlapping optical lattices \cite{meng2023atomic} or by applying the concept of synthetic dimensions \cite{Twistronics-no-twist-PRL2020}. The most challenging aspect is to engineer NH in-plane nearest-neighbor hopping in a controllable manner. To this end, we propose coupling each layer to an auxiliary optical lattice, itself coupled to an external reservoir (Fig.~\ref{fig:experimental_proposal}). This realizes the model with loss (gain) in top (bottom) layer,  as detailed in SI, Sec. S7~\cite{SI}. Its effective dynamics is analogous to that of the model with hopping imbalance,  Eq.~\eqref{eq:Ham-NH-TBG}, except for an on-site dissipative term [Eq.~(S17) in SI] that however does not alter the emergence of EMAs, as shown in SI, Sec.~S6~\cite{SI}. The feasibility of such a setup is supported by the recent experimental implementation of an analogous mechanism to induce non-Hermiticity in the Su-Schrieffer-Heeger model~\cite{NH-SSH-experiment}, indicating that it may be possible to experimentally realize NH TBG and test our theoretical predictions.

\emph{Discussion and outlook.} We here establish possibly the simplest version of NH TBG with non-Hermiticity driven by  non-reciprocal nearest-neighbor in-plane hoppings, while retaining out-of-plane tunnelings purely Hermitian. By generalizing the BM continuum model to NH TBG, we show that the non-Hermiticity yields angles at which the Hamiltonian vanishes and the bands exhibit coalescence, the EMAs, which therefore extends the flat-band region, albeit with the bands featuring finite lifetime.

 Other possible choices for NH models  of TBG may be realized by on-site dissipation or balanced gain and loss between layers. Those models also exhibit EMAs, although with a more complicated relationship between the dissipation term and the twist angle (see Sec.~S6 of  SI~\cite{SI}). As such, we argue that our minimal model describes \emph{universal} features, particularly the EMAs, that characterize  a wide class of NH bilayer Dirac systems.

The potential to realize strongly correlated phases, as for instance,  fractional quantum Hall states or unconventional superconductivity, in these NH flat band systems provides an exciting opportunity to explore interacting electronic systems in dissipative environments, particularly within the Dirac theory~\cite{Roy-TBG2019,jurivcic2024yukawa}. In future, we aim to explore the interaction effects  with particular emphasis on the EMAs, where NH effects are more pronounced. Additionally, the tunability of the imaginary part of the energy via twisting motivates the investigation of the interplay between the dissipation in these flat bands and  the magnetic field,  known to produce nontrivial features in Hermitian TBG~\cite{BM-butterfly,castro-neto-2011,koshino-PRB2012,Roy-TBG2013,Fertig-2014,choi2019electronic,Zhang2019Nearly,nuckolls2020strongly,choi2021correlation,das2021symmetry,wu2021chern,saito2021hofstadter,park2021flavour,xie2021fractional}, while in an NH setting it induces continuum bound states~\cite{Continuum-bound-states-NH}. Finally, the effects of the  effective gauge fields generated by strain~\cite{guinea2010energy,San-Jose2012non-abelian}, and  lattice buckling~\cite{NeekAmal2014,jain2016structure,Nam2017Lattice,kazmierczak2021strain} should be consequential for the band structure of the NH TBG.

The topological aspects of such NH flat bands remain largely unexplored, especially regarding topological observables; however, initial steps in this direction  have been taken  in Ref.~\cite{Huang-Yingyi-2025}.  Furthermore, we anticipate that our findings will motivate systematic study of the role of non-Hermiticity in the physics of  flat bands, not only in TBG, but also in other quantum materials. Finally,  our results should help further understand the effects of the impurities and defects in open quantum crystals featuring flat bands.

\emph{Acknowledgment.} This work is supported by the Swedish Research Council Grant No. VR 2019-04735 (V.J.), Fondecyt (Chile) Grant  No.   1230933 (V.J.). J.P.E. acknowledges support from the National Agency for Research and Development (ANID) – Scholarship Program through the Doctorado Nacional Grant No. 2024-21240412. We are grateful to Bitan Roy for the critical reading of the manuscript.

\bibliography{Bibliography}
\end{document}